\documentclass[letterpaper,twocolumn,english,aps,pra,floatfix,superscriptaddress]{revtex4}
\usepackage[T1]{fontenc}
\usepackage[latin9]{inputenc}
\usepackage{amsmath}
\usepackage{amssymb}
\usepackage{graphicx}
\usepackage{esint}

\makeatletter

\pdfpageheight\paperheight
\pdfpagewidth\paperwidth

\@ifundefined{textcolor}{}
{%
 \definecolor{BLACK}{gray}{0}
 \definecolor{WHITE}{gray}{1}
 \definecolor{RED}{rgb}{1,0,0}
 \definecolor{GREEN}{rgb}{0,1,0}
 \definecolor{BLUE}{rgb}{0,0,1}
 \definecolor{CYAN}{cmyk}{1,0,0,0}
 \definecolor{MAGENTA}{cmyk}{0,1,0,0}
 \definecolor{YELLOW}{cmyk}{0,0,1,0}
}

\renewcommand{\[}{\begin{equation}}
\renewcommand{\]}{\end{equation}}

\usepackage{babel}

\usepackage{babel}

\makeatother

\usepackage{babel}
\begin{document}
\global\long\def\avg#1{\langle#1\rangle}%

\global\long\def\p{\prime}%

\global\long\def\dg{\dagger}%

\global\long\def\ket#1{|#1\rangle}%

\global\long\def\bra#1{\langle#1|}%

\global\long\def\proj#1#2{|#1\rangle\langle#2|}%

\global\long\def\inner#1#2{\langle#1|#2\rangle}%

\global\long\def\tr{\mathrm{tr}}%

\global\long\def\pd#1#2{\frac{\partial#1}{\partial#2}}%

\global\long\def\spd#1#2{\frac{\partial^{2}#1}{\partial#2^{2}}}%

\global\long\def\der#1#2{\frac{d#1}{d#2}}%

\global\long\def\im{\imath}%

\global\long\def\S{\mathcal{S}}%

\global\long\def\A{\mathcal{A}}%

\global\long\def\F{\mathcal{F}}%

\global\long\def\E{\mathcal{E}}%

\global\long\def\M{\mathcal{M}}%

\global\long\def\As{{^{\sharp}}\hspace{-1mm}\mathcal{A}}%

\global\long\def\Fs{{^{\sharp}}\hspace{-0.7mm}\mathcal{F}}%

\global\long\def\Es{{^{\sharp}}\hspace{-0.5mm}\mathcal{E}}%

\global\long\def\EsG{{^{\sharp}}\hspace{-0.5mm}\mathcal{E}_{G}}%

\global\long\def\EsB{{^{\sharp}}\hspace{-0.5mm}\mathcal{E}_{B}}%

\global\long\def\FsG{{^{\sharp}}\hspace{-0.5mm}\F_{G}}%

\global\long\def\FsB{{^{\sharp}}\hspace{-0.5mm}\F_{B}}%

\global\long\def\Fd{{^{\sharp}}\hspace{-0.7mm}\mathcal{F}_{\delta}}%

\global\long\def\EG{\mathcal{E}_{G}}%

\global\long\def\EB{\mathcal{E}_{B}}%

\global\long\def\O{\mathcal{O}}%

\global\long\def\SgF{\S d\F}%

\global\long\def\SgEF{\S d\left(\E/\F\right)}%

\global\long\def\U{\mathcal{U}}%

\global\long\def\V{\mathcal{V}}%

\global\long\def\H{\mathbf{H}}%

\global\long\def\SO{\Pi_{\S}}%

\global\long\def\PO{\hat{\Pi}_{\S}}%

\global\long\def\SSH{\tilde{\Pi}_{\S}}%

\global\long\def\FO{\Pi_{\F}}%

\global\long\def\EO{\Upsilon_{k}}%

\global\long\def\ESH{\Omega_{k}}%

\global\long\def\HSF{\mathbf{H}_{\S\F}}%

\global\long\def\HSEF{\mathbf{H}_{\S\E/\F}}%

\global\long\def\HS{\mathbf{H}_{\S}}%

\global\long\def\ES{H_{\S}(t)}%

\global\long\def\ESo{H_{\S}(0)}%

\global\long\def\EgF{H_{\SgF} (t)}%

\global\long\def\EgE{H_{\S d\E}(t)}%

\global\long\def\EgEF{H_{\SgEF} (t)}%

\global\long\def\EF{H_{\F}(t)}%

\global\long\def\EFo{H_{\F}(0)}%

\global\long\def\ESF{H_{\S\F}(t)}%

\global\long\def\ESEF{H_{\S\E/\F}(t)}%

\global\long\def\ESSEF{H_{\tilde{\S}\S\E/\F}(t)}%

\global\long\def\EEFo{H_{\E/\F}(0)}%

\global\long\def\EEF{H_{\E/\F}(t)}%

\global\long\def\MI{I\left(\S:\F\right)}%

\global\long\def\MFF{I\left(\F:\F^{\p}\right)}%

\global\long\def\aMI{\left\langle \MI\right\rangle _{\Fs}}%

\global\long\def\BS{\Pi_{\S} }%

\global\long\def\PB{\hat{\Pi}_{\S} }%

\global\long\def\hs{\hat{s} }%

\global\long\def\Info{\mathcal{X}\left(\F\right)}%

\global\long\def\QD{\mathcal{D}\left(\Pi_{\S}:\F\right)}%

\global\long\def\QDp{\mathcal{D}\left(\PB:\F\right)}%

\global\long\def\JI{J\left(\Pi_{\S}:\F\right)}%

\global\long\def\CI{H\left(\F\left|\Pi_{\S}\right.\right)}%

\global\long\def\CIp{H\left(\F\left|\PB\right.\right)}%

\global\long\def\CS{\rho_{\F\left|s\right.}}%

\global\long\def\CSu{\tilde{\rho}_{\F\left|s\right.}}%

\global\long\def\CSp{\rho_{\F\left|\hat{s}\right.}}%

\global\long\def\CCSk{\tilde{\rho}_{k\left|\hat{s}\right.}}%

\global\long\def\CEF{H_{\F\left|s\right.}}%

\global\long\def\CEFp{H_{\F\left|\hat{s}\right.}}%

\global\long\def\psiz{\ket{\psi_{\E\left|0\right.\hspace{-0.4mm}}}}%

\global\long\def\psio{\ket{\psi_{\E\left|1\right.\hspace{-0.4mm}}}}%

\global\long\def\psiinner{\inner{\psi_{\E\left|0\right.\hspace{-0.4mm}}}{\psi_{\E\left|1\right.\hspace{-0.4mm}}}}%

\global\long\def\QDz{\boldsymbol{\delta}\left(\S:\F\right)_{\left\{  \sigma_{\S}^{z}\right\}  }}%

\global\long\def\NQD{\bar{\boldsymbol{\delta}}\left(\S:\F\right)_{\BS}}%

\global\long\def\EFS{H_{\F\left| \BS\right. }(t)}%

\global\long\def\EFSM{H_{\F\left| \left\{  \ket m\right\}  \right. }(t)}%

\global\long\def\Hol{\chi\left(\Pi_{\S}:\F\right)}%

\global\long\def\Holp{\chi(\PB:\F)}%

\global\long\def\HolF{\chi\left(\S:\Pi_{\F}\right)}%

\global\long\def\ch{\raisebox{0.5ex}{\mbox{\ensuremath{\chi}}}_{\mathrm{Pointer}}}%

\global\long\def\rhoS{\rho_{\S}(t)}%

\global\long\def\rhoSo{\rho_{\S}(0)}%

\global\long\def\rhoSF{\rho_{\S\F} (t)}%

\global\long\def\rhoSgEF{\rho_{\SgEF} (t)}%

\global\long\def\rhoSgF{\rho_{\SgF} (t)}%

\global\long\def\rhoF{\rho_{\F}(t)}%

\global\long\def\rhoFp{\rho_{\F}(\pi/2)}%

\global\long\def\LE{\Lambda_{\E}(t)}%

\global\long\def\LEc{\Lambda_{\E}^{\star}(t)}%

\global\long\def\LEij{\Lambda_{\E}^{ij}(t)}%

\global\long\def\LF{\Lambda_{\F}(t)}%

\global\long\def\LFij{\Lambda_{\F}^{ij} (t)}%

\global\long\def\LFc{\Lambda_{\F}^{\star}(t)}%

\global\long\def\LEF{\Lambda_{\E/\F} (t)}%

\global\long\def\LEFij{\Lambda_{\E/\F}^{ij}(t)}%

\global\long\def\LEFc{\Lambda_{\E/\F}^{\star}(t)}%

\global\long\def\Lkij{\Lambda_{k}^{ij}(t)}%

\global\long\def\Hb{H}%

\global\long\def\kE{\kappa_{\E}(t)}%

\global\long\def\kEF{\kappa_{\E/\F}(t)}%

\global\long\def\kF{\kappa_{\F}(t)}%

\global\long\def\ts{t=\pi/2}%

\global\long\def\QCB{\bar{\xi}_{QCB}}%

\global\long\def\CB{\xi}%

\global\long\def\mc#1{\mathcal{#1}}%

\global\long\def\MD{\lambda}%

\global\long\def\up{\uparrow}%

\global\long\def\down{\downarrow}%

\global\long\def\Cku{\rho_{k\left|\up\right.}}%

\global\long\def\Ckd{\rho_{k\left|\down\right.}}%

\global\long\def\f{\mathcal{J}}%

\global\long\def\onlinecite#1{\cite{#1}}%

\global\long\def\rs{\rho_{k\mid\hat{s}}}%

\global\long\def\ps{p_{\hat{s}}}%

\global\long\def\ppluss{p_{+\mid\hat{s}}}%

\global\long\def\pdown{p_{+\mid\downarrow}}%

\global\long\def\pup{p_{+\mid\uparrow}}%

\global\long\def\psopp{p_{-\mid\hat{s}}}%

\global\long\def\pdownopp{p_{-\mid\downarrow}}%

\global\long\def\pupopp{p_{-\mid\uparrow}}%

\global\long\def\ss{\sigma_{k\mid\hat{s}}}%

\title{Amplification, inference, and the manifestation of objective classical
information}
\author{Michael Zwolak}
\email{mpz@nist.gov}

\affiliation{Biophysical and Biomedical Measurement Group, Microsystems and Nanotechnology
Division, Physical Measurement Laboratory, National Institute of Standards
and Technology, Gaithersburg, MD, USA}
\begin{abstract}
Our everyday reality is characterized by objective information\textemdash information
that is selected and amplified by the environment that interacts with
quantum systems. Many observers can accurately infer that information
indirectly by making measurements on fragments of the environment.
The correlations between the system, $\S$, and a fragment, $\F$,
of the environment, $\E$, is often quantified by the quantum mutual
information or the Holevo quantity that bounds the classical information
about $\S$ transmittable by a quantum channel $\F$. The latter is
a quantum mutual information but of a classical-quantum state where
measurement has selected outcomes on $\S$. The measurement generically
reflects the influence of the remaining environment, $\E/\F$, but
can also reflect hypothetical questions to deduce the structure of
$\S\F$ correlations. Recently, Touil et al.~examined a different
Holevo quantity, one from a quantum-classical state (a quantum $\S$
to a measured $\F$). As shown here, this quantity upper bounds any
accessible classical information about $\S$ in $\F$ and can yield
a tighter bound than the typical Holevo quantity. When good decoherence
is present\textemdash when the remaining environment, $\E/\F$, has
effectively measured the pointer states of $\S$\textemdash this accessibility
bound is the accessible information. For the specific model of Touil
et al., the accessible information is related to the error probability
for optimal detection and, thus, has the same behavior as the quantum
Chernoff bound. The latter reflects amplification and provides a universal
approach, as well as a single-shot framework, to quantify records
of the missing, classical information about $\S$.
\end{abstract}
\maketitle
The emergence of objective, classical information from quantum systems
is due to amplification: Many pieces of the environment\textemdash e.g.,
many photons\textemdash each interact with a quantum system and acquire
an imprint of certain states, the pointer states. This is the process
by which select information becomes redundant and accessible to many
different observers. The framework, where the environment decoheres
systems and acts as a communication channel for the resulting information,
is known as quantum Darwinism \cite{ollivier_objective_2004,ollivier_environment_2005,blume-kohout_simple_2005,blume-kohout_quantum_2006,zurek_relative_2007,blume-kohout_quantum_2008,zurek_quantum_2009,zwolak_quantum_2009,paz_redundancy_2009,zwolak_redundant_2010,riedel_quantum_2010,riedel_redundant_2011,riedel_rise_2012,zwolak_complementarity_2013,zurek_quantum_2014,zwolak_amplification_2014,korbicz_objectivity_2014,brandao_generic_2015,zwolak_amplification_2016,zwolak_redundancy_2017}.
It is the pointer states that survive the interaction with the environment
and create ``copies'' of themselves from which observers can infer
the pointer state of the system. This process has been seen experimentally
in both natural \cite{unden_revealing_2019} and engineered \cite{ciampini_experimental_2018,chen_emergence_2019}
settings, and both theory and practical calculations are steadily
progressing \cite{balaneskovic_random_2015,horodecki_quantum_2015,giorgi_quantum_2015,balaneskovic_dissipation_2016,knott_generic_2018,milazzo_role_2019,campbell_collisional_2019,roszak_entanglement_2019,garcia-perez_decoherence_2020,lorenzo_anti-zeno-based_2020,ryan_quantum_2021,kicinski_decoherence_2021,korbicz_roads_2021,kwiatkowski_appearance_2021,touil_eavesdropping_2022}. 

Within this framework, one primary question concerns the information
available within an environment fragment as its size increases. This
allows one to quantify redundancy: If small fragments $\F$ of the
environment $\E$ all contain the same information about the system
$\S$, then that information is available to many observers. Given
a global state, $\rho_{\S\E}$, the accessible information
\begin{equation}
I_{\mathrm{acc}}\left(\SO\right)=\max_{\FO}I\left(\SO:\FO\right)\label{eq:AccInfo}
\end{equation}
can quantify the amount of information an observer learns about $\SO$
(a positive operator-valued measure, a POVM, on $\S$) by making a
measurement $\FO$ on only $\F$. The quantity $I\left(\SO:\FO\right)$
is the classical mutual information computed from the joint probability
distribution from outcomes of $\SO$ and $\FO$. The POVM $\SO$ has
elements $\pi_{s}$ that generate an ensemble $\left\{ \left(p_{s},\CS\right)\right\} $
of outcomes $s$ with probability $p_{s}=\tr_{\S\E}\pi_{s}\rho_{\S\E}$
and conditional states $\CS=\tr_{\S\E/\F}\pi_{s}\rho_{\S\E}/p_{s}=\tr_{\S\E/\F}\sqrt{\pi_{s}}\rho_{\S\E}\sqrt{\pi_{s}}/p_{s}$
on $\F$ (i.e., assuming the POVM acts on only $\S$ and an auxiliary
system but $\F$ is not directly affected). Allowing $\SO$ to be
arbitrary, the accessible information, Eq. \eqref{eq:AccInfo}, depicts
a situation where some auxiliary system $\A$, perhaps a special observer
or another part of the environment, has access directly only to $\S$,
makes a measurement $\SO$, and holds a record of the outcome $s$,
leaving a joint state (after tracing out the now irrelevant $\S$)
\begin{equation}
\sum_{s}p_{s}\ket s_{\A}\bra s\otimes\CS.\label{eq:AccState}
\end{equation}
An observer $\O$ then wants to predict the outcome $s$ by making
measurements only on $\F$, e.g., correlations are generated between
$\A$ and $\O$ but indirectly from separate measurements on $\S$
and $\F$, for which Eq. \eqref{eq:AccInfo} quantifies this capability.
One could then maximize the accessible information over all $\SO$
to see what quantity the observer can learn most about. This allows
one to quantify the structure of correlations between $\S$ and $\F$
induced by, e.g., a decohering interaction between them. 

Within the context of physical processes that give rise to quantum
Darwinism, $\SO$ is not arbitrary, however. For redundant information
to be present, there must be at least two records of some information,
which, when decoherence is the main interaction, will be the pointer
information. Hence, there must be an $\F$ that almost, to a degree
we want to quantify, makes a measurement of the pointer states. At
the same time, the remaining part of the environment, $\E/\F$, has
already made an effective measurement for all intents and purposes,
to a degree that we can retroactively validate. This entails that
the correlations are effectively of the form of Eq. \eqref{eq:AccState}
but with $\A=\E/\F$ or $\S$ and $\SO=\PB$ (the pointer observable),
\begin{equation}
\sum_{\hs}\ps\proj{\hs}{\hs}\otimes\CSp,\label{eq:HolState}
\end{equation}
where $\hs$ labels the pointer states (see Refs. \cite{zurek_pointer_1981,zurek_environment-induced_1982}
for discussion of pointer states). This form is a consequence of ``branching''
\cite{blume-kohout_simple_2005} and appears in the good decoherence
limit of purely decohering models, which will be extensively discussed
below. Here, it is sufficient to note that the state, Eq. \eqref{eq:HolState},
is the most relevant to quantum Darwinism. It makes little difference
if one treats the $\A$ as $\E/\F$ or as just the fully decohered,
or directly measured, $\S$, even when $\F$ is extremely large in
absolute terms \footnote{Only for ``global'' questions, where $\F$ is some sizable fraction
of the environment, does it matter. Since the environment is huge
for most problems of everyday interest, such as photon scattering,
$\F$ can be very large\textemdash even asymptotically large\textemdash without
concern for this. However, Eq. \eqref{eq:HolState} does drop exponentially
small corrections in the size of $\E/\F$ and one can not formally
take the asymptotic limit of $\F$ without first doing so in $\E$.
The degree to which asymptotic approximations work thus relies on
the balance sheet\textemdash how well records are kept in the environment
components compared to $\E$'s absolute size. Ref.~\cite{zwolak_complementarity_2013}
has dealt with retaining corrections to Eq. \eqref{eq:HolState}. }. Hereon, I will treat the auxiliary system $\A$ as if it were $\S$.

With states of the form in Eq. \eqref{eq:HolState}, the mutual information
between $\A=\S$ and $\F$ is the Holevo quantity
\begin{align}
\Holp & =H\left(\sum_{\hs}\ps\CSp\right)-\sum_{\hs}\ps H\left(\CSp\right)\label{eq:Holevo}\\
 & \equiv H_{\F}-\sum_{\hs}\ps H_{\F\left|\hs\right.},\nonumber 
\end{align}
where $H\left(\rho\right)=-\tr\rho\log_{2}\rho$ is the von Neumann
entropy for the state $\rho$. This quantity upper bounds the capacity
of $\F$ to transmit pointer state information (the variable $\hs$
is encoded in the conditional states $\CSp$). Moreover, for an important
class of interactions\textemdash purely decohering Hamiltonians with
independent environment components\textemdash the quantum Chernoff
bound determines the behavior of the optimal measurement on $\F$
to extract $\PO$ and, thus, is related to the accessible information,
Eq. \eqref{eq:AccInfo} with $\SO=\PO$. One can generalize Eq. \eqref{eq:Holevo}
by allowing one to maximize over measurements on the system, 
\begin{equation}
\chi\left(\check{\S}:\F\right)=\max_{\SO}\Hol,\label{eq:HolCheckS}
\end{equation}
where, when good decoherence has taken place, $\SO=\PB$ maximizes
the Holevo quantity \cite{zwolak_complementarity_2013}. The good
decoherence limit is when $\E/\F$ is sufficient to decohere the system
and, thus, the $\S\F$ state is exactly of the form in Eq. \eqref{eq:HolState}
\cite{zwolak_redundant_2010,zwolak_complementarity_2013}. Here, I
employ the notation $\check{\A}$ of Touil et al.~\cite{touil_eavesdropping_2022}
to indicate that the Holevo quantity is maximized over measurements
on $\A$, see also the next equation. 

Touil et al.~\cite{touil_eavesdropping_2022} examined an alternative
Holevo quantity with the measurement on the fragment side, 
\begin{equation}
\chi\left(\S:\check{\F}\right)=\max_{\FO}\HolF=\max_{\FO}\left[H_{\S}-\sum_{f}p_{f}H_{\left.\S\right|f}\right],\label{eq:HolevoF}
\end{equation}
where the maximization is over all POVMs $\FO$ and $f$ labels the
outcomes of $\FO$ and $p_{f}$ their probabilities. In that work,
they compute the quantum mutual information, the Holevo quantity in
Eq. \eqref{eq:Holevo}, and the alternative Holevo quantity in Eq.
\eqref{eq:HolevoF} for a ``{\tt c-maybe}'' model of decoherence
of $\S$ by $\E$, a model that falls into the class of purely decohering
models (see below). They analytically found $\chi\left(\S:\check{\F}\right)$
by making use of the Koashi-Winter monogamy relation \cite{koashi_monogamy_2004}
and showed all the mutual information quantities above approach the
missing information, $H_{\S}$, with a similar dependence on $\F$.

If one were to interpret this alternative Holevo quantity, Eq. \eqref{eq:HolevoF},
in the typical way, then it would bound the channel capacity of $\S$
to transmit information about (the optimal) $\FO$. One important
observation, however, is that, in the good decoherence limit\textemdash when
the $\S\F$ state is of the form in Eq. \eqref{eq:HolState}\textemdash $\chi\left(\S:\FO\right)$
lower bounds $\Holp$ for any $\FO$ by the data processing inequality
since $\PO$ is already measured on $\S$ by $\E/\F$. In this limit,
$\chi\left(\S:\check{\F}\right)$ is the actual accessible pointer
information.

For an arbitrary $\S\F$ state, however, there is no strict relation
of $\Hol$ or $\chi\left(\check{\S}:\F\right)$ with $\HolF$ or $\chi\left(\S:\check{\F}\right)$
\footnote{If $\rho_{\S\F}$ is arbitrary, then Holevo quantities with measurements
on the $\F$ side can not upper or lower bound quantities with $\S$
side measurements. For a particular state with a given inequality
between $\F$ and $\S$ side measurements, one can swap $\S$ and
$\F$ in the state $\rho_{\S\F}$\textemdash it's arbitrary after
all\textemdash and reverse the inequality.}. Instead, the inequality 
\begin{equation}
\chi\left(\S:\check{\F}\right)\ge I_{\mathrm{acc}}\left(\SO\right)\label{eq:AccessibilityBound}
\end{equation}
holds for any $\SO$. The measurement on the two sides of the inequality
is generically different\textemdash the measurement that maximizes
$\chi\left(\S:\check{\F}\right)$ is not the measurement, $\Pi_{\F}^{\star}$,
that maximizes $I\left(\SO:\FO\right)$ to get the accessible information,
Eq. \eqref{eq:AccInfo}. The proof of Eq. \eqref{eq:AccessibilityBound}
is straight forward, 
\begin{align*}
\chi\left(\S:\check{\F}\right) & =\max_{\FO}\HolF\\
 & \ge\chi\left(\S:\Pi_{\F}^{\star}\right)\\
 & =\chi\left(\M\S:\Pi_{\F}^{\star}\right)\\
 & \ge\chi\left(\SO:\Pi_{\F}^{\star}\right)\\
 & =I_{\mathrm{acc}}\left(\SO\right),
\end{align*}
where the system $\M$ is adjoined in a product state with $\rho_{\S\F}$
and a unitary on $\M\S$ makes a measurement $\SO$. The fourth line
follows from data processing.

Equation \eqref{eq:AccessibilityBound} is an accessibility bound.
Any information about $\S$ (i.e., that can be extracted by a direct
POVM on $\S$) can, at best, have $\chi\left(\S:\check{\F}\right)$
amount of shared information with $\F$. Then, as already noted, if
the good decoherence limit is reached, that bound becomes equality,
\begin{equation}
\overset{\mathbf{Good\,Decoherence}}{\chi\left(\S:\check{\F}\right)=I_{\mathrm{acc}}\left(\PO\right),}\label{eq:HolFGoodDec}
\end{equation}
for the pointer information \footnote{This follows from the form of the state in Eq. \eqref{eq:HolState}.
To determine $\chi\left(\S:\check{\F}\right)$ for this state, an
apparatus makes a measurement $\FO$ and records the outcome, leaving
a joint system-apparatus state $\sum_{\hs,f}\ps\proj{\hs}{\hs}\otimes p_{\left.f\right|\hs}\text{\ensuremath{\proj ff}}$.
This is a classical-classical state that yields, after maximizing
over $\FO$, both $\chi\left(\S:\check{\F}\right)$, Eq. \eqref{eq:HolevoF},
and the accessible information, Eq. \eqref{eq:AccInfo}.}. This makes $\chi\left(\S:\check{\F}\right)$ desirable in the context
of quantum Darwinism: It not only is a better bound on the accessible
information in the good decoherence limit\textemdash the main limit
of interest for quantum Darwinism\textemdash but it is the actual
accessible information. 

To proceed further\textemdash to compute the accessible information
and the associated redundancy\textemdash we need to specify a model
or class of models that provide the global states of interest. The
everyday photon environment has a particular structure where independent
environment components (photons) scatter off objects, acquire an imprint
of the state, and transmit that information onward, interacting little
with each other in the process \cite{joos_emergence_1985,joos_decoherence_2003,schlosshauer_decoherence_2007,riedel_quantum_2010,riedel_redundant_2011,zwolak_amplification_2014}.
This structure is captured by purely decohering Hamiltonians by independent
environment components. I will consider this general class here. Under
this evolution, the quantum Chernoff bound (QCB) provides a universal
lower bound to the accessible information and the associated redundancy.
The quantum Chernoff result is also meaningful on its own as a single-shot
result quantifying how well an individual observer (with the best
measurement apparatus) can learn the pointer state of $\S$ indirectly
from $\F$.

Pure decoherence occurs when environments select, but do not perturb,
the pointer states of $\S$. When the environment components do so
independently, the Hamiltonian is of the form 
\begin{equation}
\H=\H_{\S}+\PB\sum_{k=1}^{\Es}\EO+\sum_{k=1}^{\Es}\ESH\label{eq:Ham}
\end{equation}
 with $\left[\PB,\HS\right]=0$ and the initial state 
\begin{equation}
\rho\left(0\right)=\rho_{\S}\left(0\right)\otimes\left[\bigotimes_{k=1}^{\Es}\rho_{k}\left(0\right)\right].\label{eq:InitState}
\end{equation}
Here, $k$ specifies a component of the environment $\E$ of size
$\Es$. The operators, $\EO$ and $\Omega_{k}$, are arbitrary. This
class of models contains the {\tt c-maybe} model of Touil et al.~\cite{touil_eavesdropping_2022}.
That model has $\PB=0\cdot\proj 00+1\cdot\proj 11$ and $\exp\left[\im\EO t\right]=\sin a\proj 00+\cos a\left(\proj 01+\proj 10\right)-\sin a\proj 11$
for all $k$, where $a$ is the angle of rotation of the ``target''
environment bit after a time $t$ \footnote{Note that all the coupling frequencies (i.e., the energy scales divided
by the reduced Planck's constant) are absorbed into the definition
of the operators $\H_{\S}$, $\EO$ , and $\ESH$, while $\PB$ is
dimensionless.}. All other operators are 0. The collection of operators act similarly
to those in the controlled NOT gate. They just swap too, only a bit
more lazy, as here $a$'s any number, so it's called {\tt c-maybe}. 

Starting from the initial product state, Eq. \eqref{eq:InitState},
and evolving for some time under the Hamiltonian, Eq. \eqref{eq:Ham},
one can obtain the conditional states that appear in the Holevo quantity,
Eq. \eqref{eq:Holevo}, 
\begin{equation}
\CSp=\bigotimes_{k\in\F}\rho_{k\left|\hat{s}\right.}.\label{eq:CondFstate}
\end{equation}
Due to the structure of the evolution, these are product states over
the components of the environment fragment. However, they need not
be identically distributed (that is, they need not be fully {\tt i.i.d.}\textemdash independently
and identically distributed\textemdash states). 

The structure, Eq. \eqref{eq:CondFstate}, is a manifestation of amplification.
The pointer states $\hs$ leave an imprint on the environment components,
of which there are many. Observers intercepting those environment
components can then make a measurement to infer the pointer state.
This is the setting of quantum hypothesis testing. For instance, in
the binary case with two pointer states $\hs=0$ or $1$, one wants
to decide whether the fragment state is $\rho_{\left.\F\right|0}$
or $\rho_{\left.\F\right|1}$ with a minimum average probability of
error, $P_{e}=p_{\hs=0}\tr\Pi_{\left.\F\right|1}\rho_{\left.\F\right|0}+p_{\hs=1}\tr\Pi_{\left.\F\right|0}\rho_{\left.\F\right|1}$.
This is based on a POVM measurement, $\FO$, composed of two positive
operators $\Pi_{\left.\F\right|0}$ and $\Pi_{\left.\F\right|1}$
(with $\Pi_{\left.\F\right|0}+\Pi_{\left.\F\right|1}=\mathrm{I}$)
that indicate the occurrence of ``0'' or ``1'', respectively.
The first contribution to this average error is when the actual state
is $\rho_{\left.\F\right|0}$, with a priori probability of occurring
$p_{\hs=0}$ (where I explicitly show $\hs=0$ to connect to Eq. \eqref{eq:HolState})
but the measurement yielded the incorrect outcome $\Pi_{\left.\F\right|1}$.
Similarly for the second contribution. Moreover, when amplification
occurs, i.e., the conditional states are of the form in Eq. \eqref{eq:CondFstate},
one is specifically interested in how the error probability behaves
as the fragment size grows. This is the setting of the QCB. 

To employ the QCB, one makes use of a two-sided measurement. The first
is on $\S$, putting it into its pointer states (i.e., $\Holp$ now
provides the mutual information between $\S$ and $\F$). This reflects
the action of $\E/\F$ and is the good decoherence limit\textemdash i.e.,
$\Es\to\infty$ provided $\S$ and $\E$ have interacted for some
finite time under the evolution given by Eq. \eqref{eq:Ham} and Eq.
\eqref{eq:InitState} \footnote{This also requires that the coupling strength to the environment components
do not depend on $\Es$.}. The second is on $\F$ to access the pointer state. By Fano's inequality
\cite{cover_elements_2006,nielsen_quantum_2011}, 
\begin{equation}
\Holp\ge I_{\mathrm{acc}}\left(\PO\right)\ge H_{\S}-h\left(P_{e}\right)-P_{e}\ln\left[D-1\right],\label{eq:Fano}
\end{equation}
where $P_{e}$ is the error probability for extracting information
about a (sub)space of pointer states (of dimension $D$) from a measurement
on $\F$ \footnote{One could replace the left hand side of this inequality with $\chi\left(\check{S}:\F\right)\ge\Holp$.}.
Here, I use the binary entropy, $h\left(x\right)=-x\log_{2}x-\left(1-x\right)\log_{2}\left(1-x\right)$.
The QCB upper bound, $P_{e}^{\star}\ge P_{e}$, gives a second inequality
\begin{align}
I_{\mathrm{acc}}\left(\PO\right) & \ge H_{\S}-h\left(P_{e}\right)-P_{e}\ln\left[D-1\right]\nonumber \\
 & \ge H_{\S}-h\left(P_{e}^{\star}\right)-P_{e}^{\star}\ln\left[D-1\right],\label{eq:IntQCB}
\end{align}
which is part way to the final QCB result \cite{zwolak_amplification_2014,zwolak_amplification_2016}. 

The QCB upper bounds the error probability, $P_{e}^{\star}\ge P_{e}$,
for both the $D=2$ case \cite{audenaert_discriminating_2007,audenaert_asymptotic_2008,nussbaum_chernoff_2009}
or the $D>2$ cases \cite{li_discriminating_2016}. There is no fundamental
difference between these cases, it is only the closest two states
that determine the asymptotic decay of $P_{e}$ when $D>2$. I will
restrict to $D=2$ from hereon to make a correspondence with Touil
et al.~\cite{touil_eavesdropping_2022}. The error probability (bound)
is 
\begin{equation}
P_{e}^{\star}=\min_{0\le c\le1}p_{1}^{c}p_{2}^{1-c}\prod_{k\in\F}\tr\left[\rho_{\left.k\right|1}^{c}\rho_{\left.k\right|2}^{1-c}\right].\label{eq:PeBound}
\end{equation}
For pure $\S\E$ states in the purely decohering scenario, Eq. \eqref{eq:Ham}
and Eq. \eqref{eq:InitState}, $c$ can be any value between 0 and
1 within the generalized overlap contribution, $\tr\left[\rho_{\left.k\right|1}^{c}\rho_{\left.k\right|2}^{1-c}\right]$,
and it will give the exact overlap $\left|\inner{\psi_{\left.k\right|1}}{\psi_{\left.k\right|2}}\right|^{2}=\left|\gamma_{k}\right|^{2}$
(which is also the decoherence factor $\gamma_{k}$ squared for this
case of pure states). Touil et al.~\cite{touil_eavesdropping_2022}
consider the homogeneous case where $\gamma_{k}=\gamma$ for all $k$,
which I will also consider (see Refs.~\cite{zwolak_amplification_2014,zwolak_amplification_2016}
for inhomogeneous results). 

For pure states, therefore, only the prefactor needs optimizing over
$c$ as the generalized overlap gives $\left|\gamma\right|^{2\Fs}$
for all $0\le c\le1$ and with $\Fs$ the number of components in
$\F$. The prefactor is optimal at one of the two boundaries ($c=0$
or $c=1$), giving 
\[
P_{e}^{\star}=\min\left[p_{1},p_{2}\right]\left|\gamma\right|^{2\Fs}.
\]
I use slightly different notation here than Ref.~\cite{touil_eavesdropping_2022}
to keep the correspondence with prior work. Opposed to pure states,
for mixed $\S\E$ states within the pure decohering scenario, Eq.~\eqref{eq:Ham}
and Eq.~\eqref{eq:InitState}, the error probability (bound) is $\sqrt{p_{1}p_{2}}\prod_{k\in\F}\tr\left[\rho_{\left.k\right|1}^{1/2}\rho_{\left.k\right|2}^{1/2}\right]$
for both spin and photon models \cite{zwolak_amplification_2014,zwolak_amplification_2016}
(i.e., $c=1/2$ is optimal). Either prefactor, $\min\left[p_{1},p_{2}\right]$
or $\sqrt{p_{1}p_{2}}$, will give a bound for the pure state case.
Letting the prefactor to be just some $C$, the QCB result for pure,
homogeneous $\S\E$ is
\begin{equation}
I_{\mathrm{acc}}\left(\PO\right)\ge H_{\S}-h\left(C\left|\gamma\right|^{2\Fs}\right)\equiv\mathcal{X}_{QCB},\label{eq:QCB}
\end{equation}
where I stress that this is a classical-classical information about
random variable $\hs$ (pointer states on $\S$) with measurement
outcomes on $\F$. If we want general $\S\E$ states, but still the
pure decoherence model, Eq. \eqref{eq:Ham} and Eq. \eqref{eq:InitState},
we have exactly the same form as Eq. \eqref{eq:QCB} but the decoherence
factor (the pure state overlap) is replaced by the generalized measure
of overlap, $\tr[\rho_{\left.k\right|1}^{1/2}\rho_{\left.k\right|2}^{1/2}]$,
see Ref.~\cite{zwolak_amplification_2016} for these expressions
in terms of generic angles (between conditional states) and lengths
on the Bloch sphere for spins and Ref.~\cite{zwolak_amplification_2014}
for photons.

The QCB is a universal result. The bound Eq. \eqref{eq:PeBound} is
true for all models of pure decoherence by independent spins or the
standard photon model, all dimensions in between (qutrits, qudits,
etc.), inhomogeneous models, pure and mixed $\S\E$ states, and ones
with individual self-Hamiltonians on $\E$. The only stipulation for
Eq. \eqref{eq:PeBound} and the lower bound $H_{\S}-H\left(P_{e}^{\star}\right)$
is that one is distinguishing within a two-dimensional subspace of
$\S$ pointer states. For higher dimensional subspaces, the number
of pointer states, $D$, appears in Eq. \eqref{eq:IntQCB} and the
exponent in the decay of $P_{e}^{\star}$ requires a pair-wise minimization
of the generalized overlap over conditional states (as well as a different
prefactor outside of the exponential). 

The most important aspect of the compact form, Eq. \eqref{eq:QCB},
and its generalization to higher $D$, is that the right hand side
reflects actual, inferable information about the pointer states that
the observer can retrieve by interaction with just $\F$ in a single
shot. Moreover, while the QCB is traditionally cast as an asymptotic
result, we have not actually used any asymptotic limits to obtain
Eq. \eqref{eq:QCB}. Both of these aspects\textemdash single shot
and finite $\F$\textemdash provide a natural setting for our world,
where observers are ``agents'' within these regimes. One can then
ask questions about resources of observers (for instance, global versus
local measurements on $\F$ subcomponents \cite{kincaid_phases_nodate}
or the ability to perform coherent measurements \cite{blume-kohout_quantum_2013})
that further refine the results but do not change the fundamental
framework of single-shot, finite $\F$ inference. 

Let's return to the {\tt c-maybe} model and the Holevo quantities.
Touil et al.~\cite{touil_eavesdropping_2022} present results for
the quantum mutual information, $\chi\left(\check{\S}:\F\right)$,
and $\chi\left(\S:\check{\F}\right)$. In the good decoherence limit,
the latter two are \begin{widetext}
\begin{equation}
\chi\left(\check{\S}:\F\right)=-\frac{1}{2}\log_{2}\left[p_{1}p_{2}\left(1-\left|\gamma\right|^{2\Fs}\right)\right]-\sqrt{1-4p_{1}p_{2}\left(1-\left|\gamma\right|^{2\Fs}\right)}\mathrm{Arctanh}_{2}\left[\sqrt{1-4p_{1}p_{2}\left(1-\left|\gamma\right|^{2\Fs}\right)}\right]\label{eq:Akram17}
\end{equation}
and
\begin{equation}
\chi\left(\S:\check{\F}\right)=H_{\S}+\frac{1}{2}\log_{2}\left[p_{1}p_{2}\left|\gamma\right|^{2\Fs}\right]+\sqrt{1-4p_{1}p_{2}\left|\gamma\right|^{2\Fs}}\mathrm{Arctanh}_{2}\left[\sqrt{1-4p_{1}p_{2}\left|\gamma\right|^{2\Fs}}\right]\label{eq:Akram20}
\end{equation}
\end{widetext}in the form as they appear in their main text but using
the notation here (Eq. (17) and Eq. (20) in Ref.~\cite{touil_eavesdropping_2022}).
Rewriting these in terms of binary entropy gives
\begin{equation}
\chi\left(\check{\S}:\F\right)=h\left[\frac{1}{2}\left(1+\sqrt{1-4p_{1}p_{2}\left(1-\left|\gamma\right|^{2\Fs}\right)}\right)\right],\label{eq:Akram17RW}
\end{equation}
corresponding to the good decoherence expressions in Ref. \cite{zwolak_redundant_2010},
and
\begin{equation}
\chi\left(\S:\check{\F}\right)=H_{\S}-h\left[\frac{1}{2}\left(1+\sqrt{1-4p_{1}p_{2}\left|\gamma\right|^{2\Fs}}\right)\right].\label{eq:Akram20RW}
\end{equation}
We see that Eq. \eqref{eq:QCB} and Eq. \eqref{eq:Akram20RW} have
a similar structure. Indeed, in the good decoherence limit and for
pure conditional states, the accessible information, which is equivalent
to Eq. \eqref{eq:Akram20} or Eq. \eqref{eq:Akram20RW}, is equal
to $H_{\S}-h\left(P_{e}\right)$. Here, $P_{e}=\frac{1}{2}\left(1-\tr\left|p_{1}\rho_{\left.\F\right|1}-p_{2}\rho_{\left.\F\right|2}\right|\right)$
is the optimal error probability, which is given by the Helstrom measurement
\cite{helstrom_quantum_1976}, for single shot state discrimination
of the conditional fragment states \cite{levitin_optimal_1995,fuchs_distinguishability_1995,ban_upper_1996}.
This is not true for mixed or for higher dimensional pointer subspaces
\cite{sasaki_accessible_1999,shor_number_2000,shor_adaptive_2003,barnett_quantum_2009}.
It can be verified by a direct computation of the error probability
from the optimal measurement for the pure conditional states. For
$\CSp$ pure, the trace distance in the Helstrom expression just requires
diagonalizing an operator in a two-dimensional subspace, giving 
\begin{equation}
P_{e}=\frac{1}{2}\left(1-\sqrt{1-4p_{1}p_{2}\left|\gamma\right|^{2\Fs}}\right)\label{eq:OptPe}
\end{equation}
(this readily generalizes to the inhomogeneous case: The factor $\left|\gamma\right|^{2\Fs}$
just needs to be replaced by $\prod_{k\in\F}\left|\gamma_{k}\right|^{2}$).
This result makes no use of the fact that the environment components
were spins, and thus it is directly applicable to (pure state) photon
scattering off an object in a two dimensional superposition, more
directly supporting the connection discussed in Touil et al.~\cite{touil_eavesdropping_2022}
and extending it to $\chi\left(\S:\check{\F}\right)$ in the good
decoherence limit \footnote{Moreover, as with the QCB result, the form of the accessible information
for pure $\S\E$ states, $H_{\S}-h\left(P_{e}\right)$ , with the
optimal $P_{e}$ from Eq. \eqref{eq:OptPe} holds regardless of the
environment components. They can be spins, qudits, or photons. Furthermore,
the connection with hypothesis testing allows for even more general
statements about models that are not purely decohering. For instance,
for projection-valued measurements and pure $\S\E$ states, one obtains
the same accessible information, $H_{\S}-h\left(P_{e}\right)$, but
the error probability just has the overlap between the conditional
fragment states, which can behave in a manner that is not exponentially
decaying with $\Fs$.}. 

While specific to the case of $D=2$ and pure $\S\E$ states evolving
under Eq. \eqref{eq:Ham} and Eq. \eqref{eq:InitState}, the connection
provides a window into the behavior of different ways to quantify
correlations. The alternate Holevo quantity, $\chi\left(\S:\check{\F}\right)$,
becomes the inferable information in this specific setting. However,
inferable information has a universal form that goes beyond this specific
setting of dimensionality and purity. 

\emph{Redundancy}. The decay to the classical plateau\textemdash the
missing information $H_{\S}$ about the system\textemdash for the
quantities in Eq. \eqref{eq:Akram17RW}, Eq. \eqref{eq:Akram20RW},
and Eq. \eqref{eq:QCB} all are controlled by the $\F$-induced decoherence
factor, $\gamma^{2\Fs}$. Ultimately, though, we are interested in
the redundancy of information. This requires introducing a control,
the information deficit $\delta$, that reflects the fact that one
can not generally obtain perfect knowledge from a finite-size fragment
$\F$. This is typically taken as 
\begin{equation}
\Info\ge H_{\S}\left(1-\delta\right),\label{eq:RedThreshold}
\end{equation}
where $\Info$ is some mutual information (quantum mutual information,
Holevo, accessible information, etc.). This is the form I will employ
here. However, both the form of the QCB and the form of $\chi\left(\S:\check{\F}\right)$
(in the good decoherence limit) suggest employing the information
deficit as an entropic quantity when thresholding entropic measures
of information,

\[
\Info\ge H_{\S}-H\left[\delta\right].
\]
This allows $\delta$ to be a factor reflecting distinguishability
of conditional states and allows for non-asymptotic computations to
proceed for the redundancy (it removes the transcendental form of
the equations). I will not use this in what follows. 

The approach to the plateau and the redundancy (to within $\delta$)
have simple asymptotic results regardless of quantity used to compute
them. The decay exponent to the plateau, $\xi$, of some information
theoretic quantity $\Info$, such as Eq. \eqref{eq:Akram17RW}, Eq.
\eqref{eq:Akram20RW}, or Eq. \eqref{eq:QCB}, is
\begin{equation}
\xi=-\lim_{\Fs\to\infty}\frac{1}{\Fs}\ln\left[H_{\S}-\Info\right].\label{eq:AsymptoticDecay}
\end{equation}
For the pure, homogeneous {\tt c-maybe} model, all three decay to
the plateau with exponent 
\begin{equation}
\xi=-\ln\left|\gamma\right|^{2}.\label{eq:UniDecay}
\end{equation}
That is universality in a nutshell. Moreover, the exponent is the
leading order of the redundancy, 
\begin{equation}
R_{\delta}\simeq\Es\frac{\xi}{\ln1/\delta}=\Es\frac{\ln\left|\gamma\right|^{2}}{\ln\delta}.\label{eq:Red}
\end{equation}
This is the essence of the QCB: The exponent\textemdash the quantum
Chernoff information, $\xi_{QCB}$, or its inhomogeneous counterpart,
$\bar{\xi}_{QCB}$\textemdash controls the redundancy, see Refs.~\cite{zwolak_amplification_2014,zwolak_amplification_2016}
for additional discussion and results. For the pure {\tt c-maybe}
model, this exponent is the same whether using Eq. \eqref{eq:Akram17RW},
Eq. \eqref{eq:Akram20RW}, or Eq. \eqref{eq:QCB}. The quantum mutual
information also yields the same decay and redundancy in the good
decoherence limit, as it is the same as $\chi\left(\check{\S}:\F\right)$
from Eq. \eqref{eq:Akram17RW} \footnote{In order to apply Eq. \eqref{eq:AsymptoticDecay} for the quantum
mutual information, one needs $\Es\to\infty$. As already mentioned
in a prior footnote, though, this will entail good decoherence provided
some finite interaction between $\S$ and $\E$ components has taken
place. }. In other words, all the information theoretic quantities provide
the same decay and redundancy, which the asymptotic calculation, Eq.
\eqref{eq:AsymptoticDecay}, makes apparent in a non-empirical manner.
\begin{figure}[t]
\begin{centering}
\includegraphics[width=1\columnwidth]{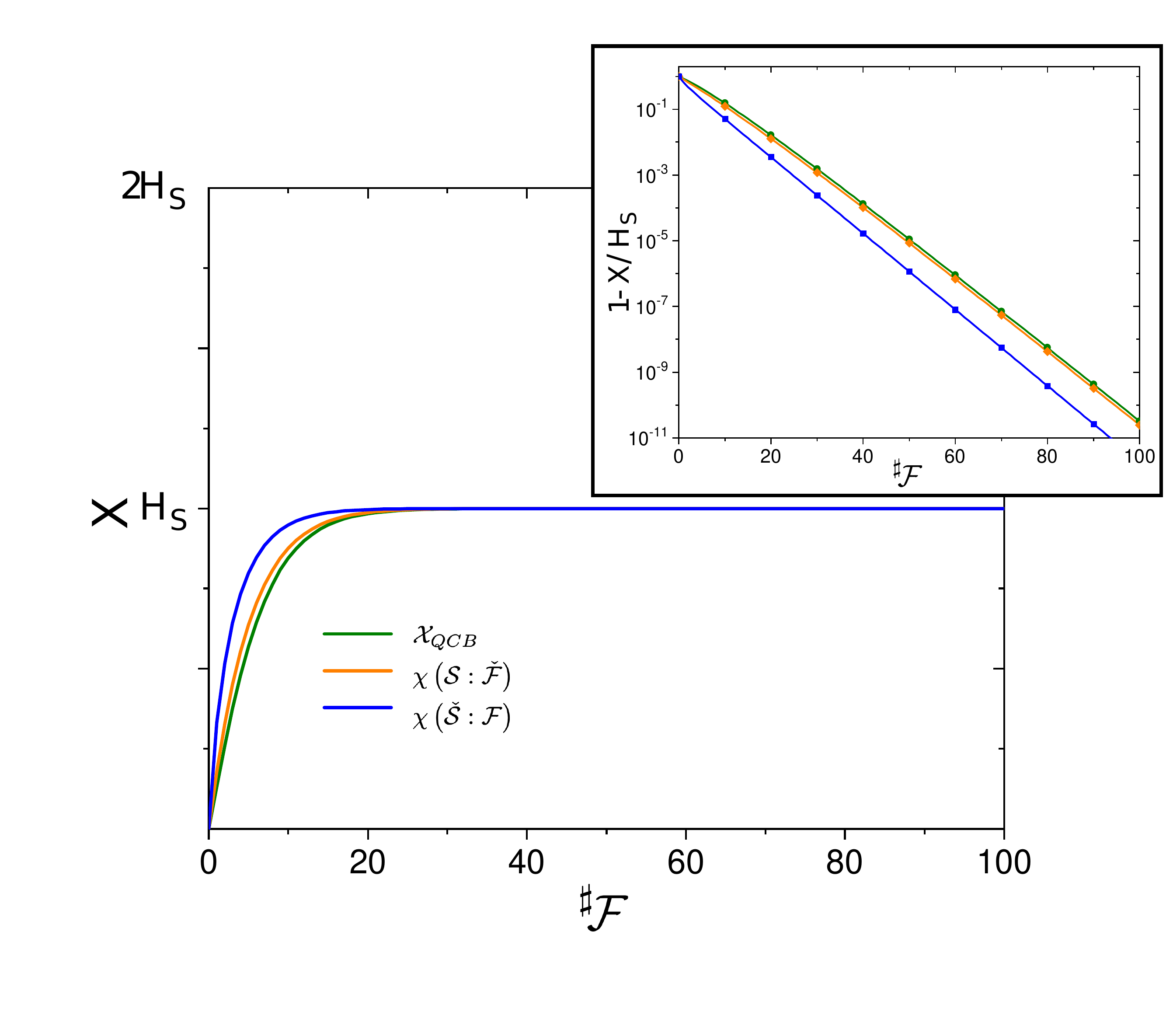}
\par\end{centering}
\caption{\textbf{Approach to the plateau.} Information measures $\mathcal{X}$
versus fragment size $\protect\Fs$ for $p_{1}=1/4$ and $\gamma=7/8$.
All three quantities, $\mathcal{X}=\mathcal{X}_{QCB}$ (green line),
$\chi\left(\protect\S:\check{\protect\F}\right)$ (orange line), and
$\chi\left(\check{\protect\S}:\protect\F\right)$ (blue line), rapidly
rise to the classical plateau, $H_{\protect\S}$, as the fragment
size $\protect\Fs$ increases. The quantum mutual information, $\protect\MI$
(not shown), is equivalent to $\chi\left(\check{\protect\S}:\protect\F\right)$
when good decoherence is present. The QCB result, $\mathcal{X}_{QCB}$,
lower bounds the other two, but is close to $\chi\left(\protect\S:\check{\protect\F}\right)$.
The inset shows the decay to the plateau. All three measures decay
with the same exponent. The $\chi\left(\check{\protect\S}:\protect\F\right)$
does, though, deviate from the other two quantities, as the latter
two have a prefactor that depends on $\protect\Fs$ (both with the
same functional form). This offset does not influence the redundancy
asymptotically (i.e., as a relative correction, it itself decays).
\label{fig:PlateauApproach}}
\end{figure}

Figure \ref{fig:PlateauApproach} shows the approach to the plateau
for the three information measures. The quantity $\chi\left(\check{\S}:\F\right)$
is a weaker bound to the accessible information. Yet, the separation
between the decay curves is unimportant for passing the threshold
in Eq. \eqref{eq:RedThreshold}: $\chi\left(\check{\S}:\F\right)$
passes it sooner than the other quantities, but this only gives a
relative correction to Eq. \eqref{eq:Red} that goes to zero asymptotically
($\Fs$ and $-\ln\delta$ have to simultaneously go to infinity),
albeit weakly as $1/\ln\delta$ \footnote{To clarify this statement, let $R_{\delta}=R_{\delta}^{\circ}+R_{\delta}^{\p}$,
with $R_{\delta}^{\circ}$ from the right hand side of Eq. \eqref{eq:Red}
and $R_{\delta}^{\p}$ the corrections. The relative correction, $R_{\delta}^{\p}/R_{\delta}^{\circ}$
decays as $1/\ln\delta$ for $\chi\left(\check{\S}:\F\right)$ and
as $\ln\left(\ln1/\delta\right)/\ln\delta$ for $\chi\left(\S:\check{\F}\right)$
and $\mathcal{X}_{QCB}$ as $\delta\to\infty$. In other words, $R_{\delta}^{\p}\sim1/\left(\ln\delta\right)^{2}$
asymptotically. The very weak prefactor, $\ln\left(\ln1/\delta\right)$,
for the latter two cases is due to the presence of $\Fs$ in the prefactor
in Eqs. \eqref{eq:HolFDecay} and \eqref{eq:QCBDecay}.}. The leading order contribution to the decay for $\chi\left(\check{\S}:\F\right)$
is
\begin{equation}
\frac{p_{1}p_{2}\log_{2}\frac{p_{2}}{p_{1}}}{p_{2}-p_{1}}\left|\gamma\right|^{2\Fs}\label{eq:HolSDecay}
\end{equation}
or with a prefactor of $1/2\ln2$ when $p_{1}=p_{2}=1/2$. For $\chi\left(\S:\check{\F}\right)$,
the decay is 
\begin{equation}
p_{1}p_{2}\log_{2}\left[\frac{e}{p_{1}p_{2}}\left|\gamma\right|^{-2\Fs}\right]\left|\gamma\right|^{2\Fs}\label{eq:HolFDecay}
\end{equation}
and, for the QCB result,
\begin{equation}
C\log_{2}\left[\frac{e}{C}\left|\gamma\right|^{-2\Fs}\right]\left|\gamma\right|^{2\Fs}\label{eq:QCBDecay}
\end{equation}
with $C=\min\left[p_{1},p_{2}\right]$ or $\sqrt{p_{1}p_{2}}$ depending
on whether we take the pure state result or generically take the mixed
state bound. These forms show the same exponential decay but the latter
two have a weak dependence of the prefactor on $\Fs$. 

To summarize, quantum Darwinism clarifies the role of the proliferation
of information in the quantum-to-classical transition. Here, I examined
the quantity introduced by Touil et al.~\cite{touil_eavesdropping_2022},
$\chi\left(\S:\check{\F}\right)$, where an (optimal) measurement
is made on the fragment, reminiscent of the quantum Chernoff bound.
It provides an appealing approach to finding the redundancy of information,
as it is an accessibility bound that becomes the accessible information
in the limit of good decoherence. For the special case of a pure $\S\E$
state, the accessible information is directly related to the optimal
error probability for distinguishing conditional states on the environment
(i.e., hypothesis testing or inference), of which an exact expression
(including the prefactor) can be computed. Moreover, this connection
immediately generalizes the result to any pure, $D=2$ model (spin
environments, qudit environments, photon environments, etc.)~and
to inhomogeneous environments (including ones with self-Hamiltonians,
as in Eq. \eqref{eq:Ham}). That decay, as expected, has the same
exponent as the QCB, as the QCB promises (and only promises) to yield
the right asymptotic decay, not the prefactor. Asymptotic analysis
provides a non-empirical way to show that all quantities give the
same redundancy\textemdash due to the same exponent\textemdash to
leading order (and that corrections are small) and makes universality
of the plateau approach manifest. Since the QCB applies more generally,
its universal bound should further help shed light on future results
that yield exact entropic quantities or alternative bounds. Its importance\textemdash the
QCB's importance\textemdash goes beyond this, though, as it provides
a single shot, finite $\F$ framework for understanding how we observers
learn in a quantum Universe. 

I would like to thank W. H. Zurek for inspiration, many years of entertaining
and enlightening discussions, and moneywine bets on various scientific
topics. I would also like to thank J. Elenewski and A. Touil for helpful
comments on this manuscript.

\end{document}